# Aging the Cu-doped Bi$_2$Te$_3$ crystals for the topological transport and its atomic tunneling-clustering dynamics

Taishi Chen[1], Qian Chen[2], Koen Schouteden[3], Wenkai Huang[1], Xuefeng Wang[4], Zhe Li[3], Feng Miao[1], Xinran Wang[4], Zhaoguo Li[1], Bo Zhao[1], Shaochun Li[1], Fengqi Song[1,*], Jinlan Wang[2,*], Baigeng Wang[1,*], Chris Van Haesendonck[3], Guanghou Wang[1]

[1] National Laboratory of Solid State Microstructures and department of Physics, Nanjing University, 210093, Nanjing, P. R. China

[2] Department of Physics, Southeast University, 211189, Nanjing, P. R. China

[3] Laboratorium voor Vaste stoffysica en Magnetisme, K. U. Leuven, BE-3001, Belgium

[4] School of Electronic Science and Engineering, Nanjing University, 210093, Nanjing, P. R. China

---

[*] Corresponding authors. F. S (songfengqi@nju.edu.cn), J.W. (jlwang@seu.edu.cn), B. W. (bgwang@nju.edu.cn). Fax: +86-25-83595535. The first 4 authors contributed equally.

Enormous efforts have been devoted to suppressing the bulk conduction in the study of three-dimensional topological insulators (TIs) of the $Bi_2Se_3$ families [1-4] because the topological surface state (TSS) transport is commonly hindered by the bulk electrons due to the imperfect electron-hole hybridization, intrinsic Se/Te vacancies in real TI samples [5-10]. For example, thickness reduction has been used as an effective strategy to enhance the SS transport [8, 11]. But even with an insulating mother crystal, the exfoliated samples may still exhibit a diffusive transport since the samples suffer from unexpected contaminations [12]. In addition, there always exists some local roughness, leading to non-uniformity in ultrathin TI layers prepared by molecular beam epitaxy [13]. Both aspects make proper preparation of ultrathin TI layers laborious [4]. Another approach is to tune the TI samples (Fermi level) to charge neutrality by the compensation of doped charges. Using ternary ($Bi_2Te_2Se$) crystals has been successful since stoichiometric $Bi_2Te_3$ and $Bi_2Se_3$ contribute opposite types of carriers [6, 14]. However, the Fermi level is often located near the conduction band. Such close spacing may even couple the bulk and surface electrons with the result of trivial transport up to very large thickness [15]. Moreover, this sometimes leads to a topologically-trivial crystal due to a change of spin orbit interaction [16]. Intentional doping with the elements such as Ca, Sn have also been tried [17-18]. Although 0.1% of dopant is able to suppress the bulk carrier concentration, a serious concern is that the Fermi level will further shift due to the migration of the dopant atoms after the optimized TSS has been achieved [17-18]. Here we introduce another strategy, i.e. aging, after which the dopant atoms become stabilized. We have successfully observed the

TSS-induced transport in bulk $(Cu_{0.1}Bi_{0.9})_2Te_{3.06}$ crystals. An atomic tunneling-clustering picture is proposed to interpret the Cu migration across a 0.57eV-high barrier.

All our crystals are prepared by the high-temperature sintering method. An aging process is then followed with the resultant samples described in Table 1 and supplementary materials. **Figure 1** presents their electrical transport data. The relative magnetoconductance (MC) is plotted in Fig 1(a). A parabolic field dependence is found for Sample 1 and 2, while a tip-shaped MC feature near zero fields is found for Sample 3 and 4. The parabolic MC curve is typical of normal metallic transport. The MC curves for Sample 3 and 4 can be attributed to weak antilocalization (WAL), which is regarded as the signature of TSS transport [19-25]. Therefore the above WAL dominance tends to reflect the enhanced TSS transport in aged samples as demonstrated below. Let's see Fig 1(b) for their temperature-dependent resistivity. The samples exhibit a positive temperature-resistance (RT) dependence before aging (sample 1), while it gradually turns to negative (Sample 4). In Sample 4, the low-temperature resistivity reaches over 100 mΩcm. Fig. 1(c) also presents the temperature-dependent mobility of the samples, which drops by four orders of magnitude after intense aging. This suppression allows for the obvious TSS transport in bulk crystals.

The angular MC dependence demonstrates the WAL character of the 2D TSS. **Figure 2**(a) presents the magnetoresistance (MR) of Sample 4 measured along different directions at 2 K, where the WAL tips with increasing angles as marked by the blue shadow in the inset. However, all low-field MR curves collapse on each other after normalizing the field to its perpendicular component in Fig 2(b). This indicates that the observed WAL arises from a 2D electronic state [6, 9, 26-27]. The 2D WAL indicates the presence of a coherent SS confined within a depth that is much less than the dephasing length. Considering a dephasing length of over 300 nm and a sample thickness of 30 μm, this confirms the TSS-mediated transport in our TI crystals [28]. The 2D TSS transport is further confirmed by more data analysis. Fig 2(c) shows the WAL features at different temperatures, where all the curves are nicely fitted by the Hikami-Larkin-Nagaoka equation [19, 25, 29]. The fittings provide the values of dephasing lengths and channel numbers. The dephasing length decreases with the increasing temperature as shown by Fig 2(d). An ln-ln fitting is carried out to identify the temperature scaling [30], giving an exponential constant of 0.55±0.06. It is typical of 2D electronic interference [19, 31]. All the evidence supports the TSS transport in our aged crystals.

Additional evidence is provided by scanning tunneling microscopy (STM). An STM image of a triangular defect in Sample 4 is presented in Fig 2(e). The corresponding map of the local density of states in Fig 2(f) reveals the presence of complex wave patterns near the boundaries. Such patterns can be interpreted as standing waves due to 2D electronic interference, of TSS electrons, which are scattered at the defect boundaries. Figure 2(g) presents d$I$/d$V$ curves of various locations on Sample 4. Their features are in agreement with previous data [13]. The Fermi level is located within the bandgap at 0 meV, i.e., 60 meV lower than the bottom of bulk conduction band (BCB) and 100 meV higher than the top of bulk valence band (BVB). The bandgap is consistent with thermal activation based calculations (Fig 1(b)). This confirms our material is with an insulating bandgap.

The parallel conduction of the bulk state and TSS should lead to a bent Hall curve, as indeed observed in the inset of Fig 2(a) [6, 14]. Two-channel analysis reveals that the ratio of the surface conductance to overall conductance is 3.3%. The charge concentration and mobility of the TSS are determined $2.0 \times 10^{12} cm^{-2}$ and $1400 cm^2 V^{-1} S^{-1}$ respectively. The mobility of bulk electrons is $2.4\ cm^2 V^{-1} S^{-1}$. We also obtain some 100nm-thick flakes, which low-temperature transport are dominated by the TSSs [32]. We find that a few months of further exposure to the air don't make significant influence on the samples. Therefore, such optimized materials are useful for future TI applications.

The question then arises what happens and leads to the mobility suppression of the bulk electronic states during the aging. We check the possible occurrence of oxidation by X-ray photoemission spectroscopy (XPS). The XPS data are collected while etching the sample with an Ar ion beam. The analysis of the XPS data reveals that all the O signals are confined to less than 20 nm from the surface. Oxidation cannot induce disorder in the bulk of the crystals with a thickness of 30 μm. Oxidation is therefore excluded as a possible cause of the observed mobility suppression. H2O absorption can also be excluded since it is again not possible to account for a suppression of the complete crystal. Such conjecture is certainly reasonable since the crystals are sealed in vacuum during the aging.

**Figure 3(a)** presents an atomically resolved STM image of an atomically flat terrace of a freshly cleaved (ex situ) pristine $Bi_2Te_3$ flake, which is free of defects and adsorbates. After the doping, two features are found (Figs. 3(b), (c) and (d)): nanometer-size islands (bright colored in Fig 3(e)) and triangularly shaped defects (dark colored in Fig 3(f)). The amount of Cu dopants is found to scale with the amount of triangularly shaped defects. The islands are also observed on samples that are cleaved in situ (see Figs 3(b) and (d)). Such islands are absent on pristine $Bi_2Te_3$. We therefore conclude that both the islands and the triangularly shaped defects are induced by the doping. The triangular defects can be attributed to the Cu dopants inside the quintuple layers (QL) [33] and the bright islands are Cu clusters. The STM image of a typical $Cu_3$ cluster is presented in Fig 3(e). The height profile in Fig 3(d) reveals the presence of a 1nm-high terrace and a 0.4nm-high island, which is typical for atomic-size Cu islands. Cu atoms can reside in three different positions in the crystals, i.e. inside the QLs (position I in **Figure 4**), in between the QLs (position II) and at the interfaces of the two positions. Cleavage always occurs between QLs and therefore the observed islands can be assigned to Cu clusters of position II. Changes of the two features can then be interpreted in terms of the dynamics of the Cu atoms in the $Bi_2Te_3$ crystals. Note that Fig. 3(b) is obtained for a sample without aging, while Figs. 3(c) and (d) are obtained for an aged sample. We see that the bright islands become more abundant after the aging. The measurements thus indicates that the Cu atoms, which are initially at position I, migrate to position II during the aging. Subsequently the Cu atoms then aggregate to form the Cu clusters as observed in Figs.

3(b) and (c).

The structural analysis by high resolution transmission electron microscopy (HRTEM) demonstrates the degrading of the crystalline quality after aging [34]. Fig. 3(g) presents a typical morphology of the pristine $Bi_2Te_3$ and of the $(Cu_{0.1}Bi_{0.9})_2Te_{3.06}$ before aging. The crystal flakes can be very large, up to over 200 x 200nm$^2$. Taking the HRTEM images along the edge of the flake, we are able to observe some parallel fringes with a lattice spacing of 1 nm. These are the (003) fringes of the $Bi_2Te_3$ QLs. Moreover, we can also observe the ideal hexagonal lattice in other regions of the flake in the Cs-corrected HRTEM images[35-36]. The HRTEM analysis thus reveals very good crystalline structure of our samples before aging. After aging we only observe much smaller crystalline domains, as illustrated in Fig. 3(g). The typical dimensions of the domains are around 5 nm, 10-20 nm in some cases. The HRTEM results provide evidence for the decay of the crystalline order during the aging. The observed poor crystalline order will naturally accounts for the observed pronounced suppression of the bulk mobility.

The first-principles density functional theory (DFT) is employed to describe the atomic dynamics of Cu dopants in $Bi_2Te_3$. We use $E_f$ (the formation energy) to evaluate the stability of different Cu dopant positions in $Bi_2Te_3$ [37], including position I, position II and at the interface (the "transient" state). At first, we check the positions inside the QLs (position I), where Cu atoms occupy the substitutional positions (Bi or Te) and interstitial positions, respectively. The interstitial positions in Te layer are found to be energetically favored, as shown in **Figure 4**. Next, a second calculation is carried out for position II, which shows that Cu atoms prefer to adsorb on one side of the gap, in a hollow site of three Te atoms. And the formation energy here is very close to those atoms inside the QLs. Finally, the migration path of the Cu atom diffusion from the QLs to the gap is calculated. A significant reaction barrier as high as 0.57eV is located, which limits the direct exchange of Cu atoms in between them despite of their similar binding energies. While, as the Cu atom is rather small as compared to the QL gap [38-41], the possibility of the formation of Cu clusters also need consider. The result reveals that the formation of $Cu_3$ clusters in between the QLs is preferred with a 0.16 eV/Cu atom, larger binding energy than for an individual dopant atom. On the other hand, the system with $Cu_3$ clusters exhibits a total energy similar to the energy of the system with separate Cu atoms inside the QLs.

A clear "tunneling-clustering" picture with a diffusion barrier is revealed, where Cu atoms can migrate freely both in and between the QLs, while they have to overcome a 0.57eV high barrier when crossing the interface. During the long aging period, the dopant atoms diffuse inside and between the QLs and frequently challenge the barriers. As the formation of Cu clusters between the QLs determines the final direction of the atomic dynamics toward the QL gaps, Cu atoms in the QLs will gradually climb over the barrier and form clusters between the QLs, as seen in the experiment [42]. The crystalline quality of the QL will degrade due to the migration of a vast amount of Cu atoms during the aging, finally leading to smaller crystalline domains and strongly suppressed mobilities of the bulk electrons.

**In conclusion,** the TSS transport is observed in bulk crystals of aged

$(Bi_{0.9}Cu_{0.1})Te_{3.06}$ due to the suppressed bulk mobility by four orders of magnitude. This results in an optimized and band-insulating TI crystals. The STM/HRTEM and DFT calculations reveal the atomic dynamics during the aging, when the dopant atoms challenge through a 0.57eV-high diffusion barrier and form Cu clusters. Such dopant migrations degrade the crystalline quality of the crystal and leads to the final dominance of the TSS transport.

**Acknowledgement**

We thank the National Key Projects for Basic Research of China (grant numbers: 2013CB922103, 2011CB922103, 2010CB923401, 2011CB302004 and 2013CBA01600), the National Natural Science Foundation of China (grant numbers: 11023002, 11134005, 60825402, 61176088, 11075076, 21173040, 61261160499 and 11274154), NSF of Jiangsu province (No. BK2011592, BK20130016, BK20130054, BK2012322 and BK2012302), the PAPD project and the Fundamental Research Funds for the Central Universities for financially supporting the work. Helpful assistance from Nanofabrication and Characterization Center at Physics College of Nanjing University, Prof. Mingxiang Xu and Dr Longbing He at Southeastern University, Prof. Yongqing Li at institute of Physics in Beijing, Dr. Li Pi and Prof. Yuheng Zhang at High Magnetic Field Laboratory CAS are acknowledged. JW thanks the computational resources at SEU and National Supercomputing Center in Tianjin. K. Schouteden is a postdoctoral researcher of the Research Foundation − Flanders (FWO, Belgium). Z. Li thanks the China Scholarship Council for financial support (grant number 2011624021).

[28] One may question the existence of 2D electron gas. Here we exclude such possiblity since the 2D electron gas relies on the local environment. e.g. it should be different in different positions in a sample. Our observations of 2D transport are common in the low-mobility samples`, therefore`, we suggest its TSS origin.

[30] The obtained channel number α is shown in the right inset of Fig 3(d)`, nearly 3 at low temperatures. It decays to 2.3 at higher temperatures due to the possible coupling. The value of around 3 indicates the presence of 6 TSS channels`, i.e. 6 surfaces. This can be understood when considering the left inset of Fig 3(d)`, which is an SEM image of the sample's sidewall. We observe the presence of some crack-like features on the side wall`, which can also be observed for other layered crystals. Such features may be related to the mechanical cleavage and expose some more surfaces.

[32] The condutance of the 30um flake consists both the surface and bulk contribution as formulated. $G=G_s+G_b=G_s+H g_b$`, where $G_s$ is the sheet surface conductance`, $g_b$ is the sheet bulk conductivity and H is the thickness. The conductance of a 160nm-thick sample is close to $G_s$. This demonstrates the dominance of the TSS.

**Figure Captions:**

Figure 1. **The comparison of the four samples with different aging from the aspects of 2K magnetoresistance (MR) (magnetoconductance(MC)) (a), temperature-dependent resistivity (b) and mobility (c).** In (a), The MC curves for the Sample 1 and 2 are plotted on the right axis and those for Sample 3 and 4 are plotted on the left axis. (b) The temperature-dependent resistivity of our samples.

Figure 2. **Experimental evidence on the 2D Topological Surface Position In Sample-4.** (a) The angle-dependent MR curves of Sample-4. The inset is its Hall curve. (b) The MR curves plotted against the perpendicular component of the magnetic field. The inset is the initial MR curves. (c) Temperature dependent of the relative MC of the sample-4 at different temperatures that can be well fit by the equation. (d) The scaling of dephasing lengths upon the temperatures. The right inset is the temperature-dependent α. The left inset is the SEM image. (e) and (f) are respectively STM images and spectrum image around a defect. (g) is the dI/dV curve on sample 4,. The lack dash line indicates the position of the Fermi level. BVB and BCB mark the positions of valance and conduction band respectively, relatively -0.10eV and 0.06eV higher than the Fermi level.

Figure 3. **Defects characterizations by STM and TEM.** (a) is the STM image of a pristine $Bi_2Te_3$ crystal with a scale bar of 1nm. (b) and (c) are the STM results from two flakes extracted from the unaged and the aged sample of $(Cu_{0.1}Bi_{0.9})_2Te_{3.06}$, corresponding to the Sample-1 and Sample-4, respectively. (d) is the large-scale

scanning of Sample 4, confirming the dominance of the light islands. A linear plot of the Sample-4 is shown as the inset of (d), indicating the white islands' heights of about 0.4nm, typically for Cu clusters. (e) the atomic-resolution image of the smallest light islands, which is obviously a 3-atom cluster, $Cu_3$ clusters. (f) the high-resolution image of the dark triangle, which depth is 0.04nm. (g) HRTEM image of the unaged samples, where the 1nm-period fringes appear along the edge and the hexagonal lattices appear in the whole flake. Its model is illustrated by its inset.   (h) Typical HRTEM image of Sample 4, where the crystals are much smaller.

**Figure 4. modeling the aging process.** The 4 images illustrate the formation energies of $Cu_3$ clusters inside QLs, individual Cu inside QLs (position I), individual Cu between QLs (in gap, position II) and $Cu_3$ clusters between QLs (in gap) respectively. In each image, the top part is the side view and the bottom greyer part is the top view. The formation energies are marked on the top.

| Sample (Name) | Size (mm×mm×μm) | Resistivity at 2K (mΩ cm) | Mobility ($cm^2$/Vs) | Resistivity at 300K (mΩ cm) |
|---|---|---|---|---|
| Sample-1 | 2.3×1.1×20 | 0.3 | 1980 | 1.8 |
| Sample-2 | 2.4×1.2×50 | 8.3 | 925.1 | 7.6 |
| Sample-3 | 2.2×1.0×32 | 48.0 | 1.8 | 9.8 |
| Sample-4 | 4.1×1.0×40 | 72.4 | 2.6 | 10.2 |

Table 1. the parameters of the $(Cu_{0.1}Bi_{0.9})2Te_{3.06}$ samples. Sample 1 is free of aging. Sample 3 and 4 are aged for 600 days. Sample 2 are with half aging extent of them.

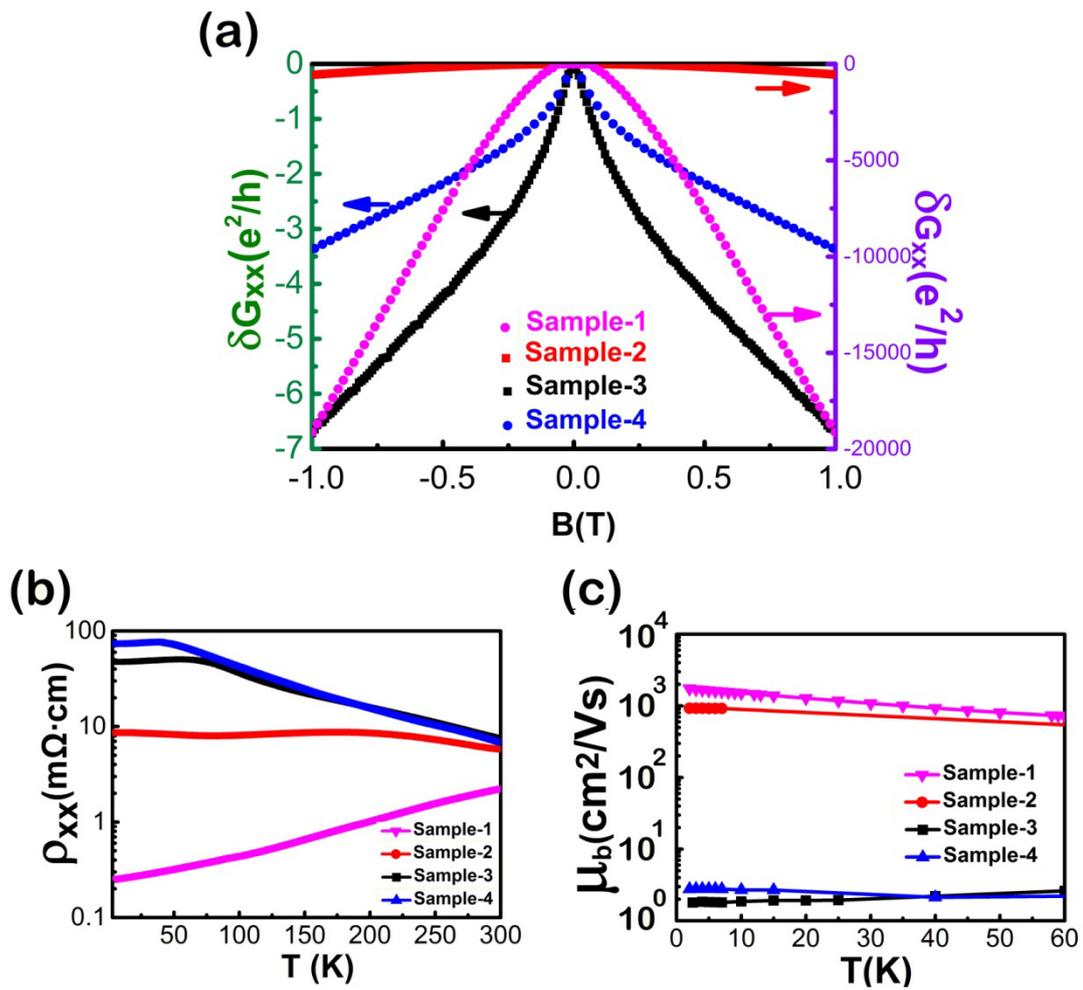

Figure 1. The comparison of the four samples with different aging from the aspects of 2K magnetoresistance (MR) (magnetoconductance(MC))

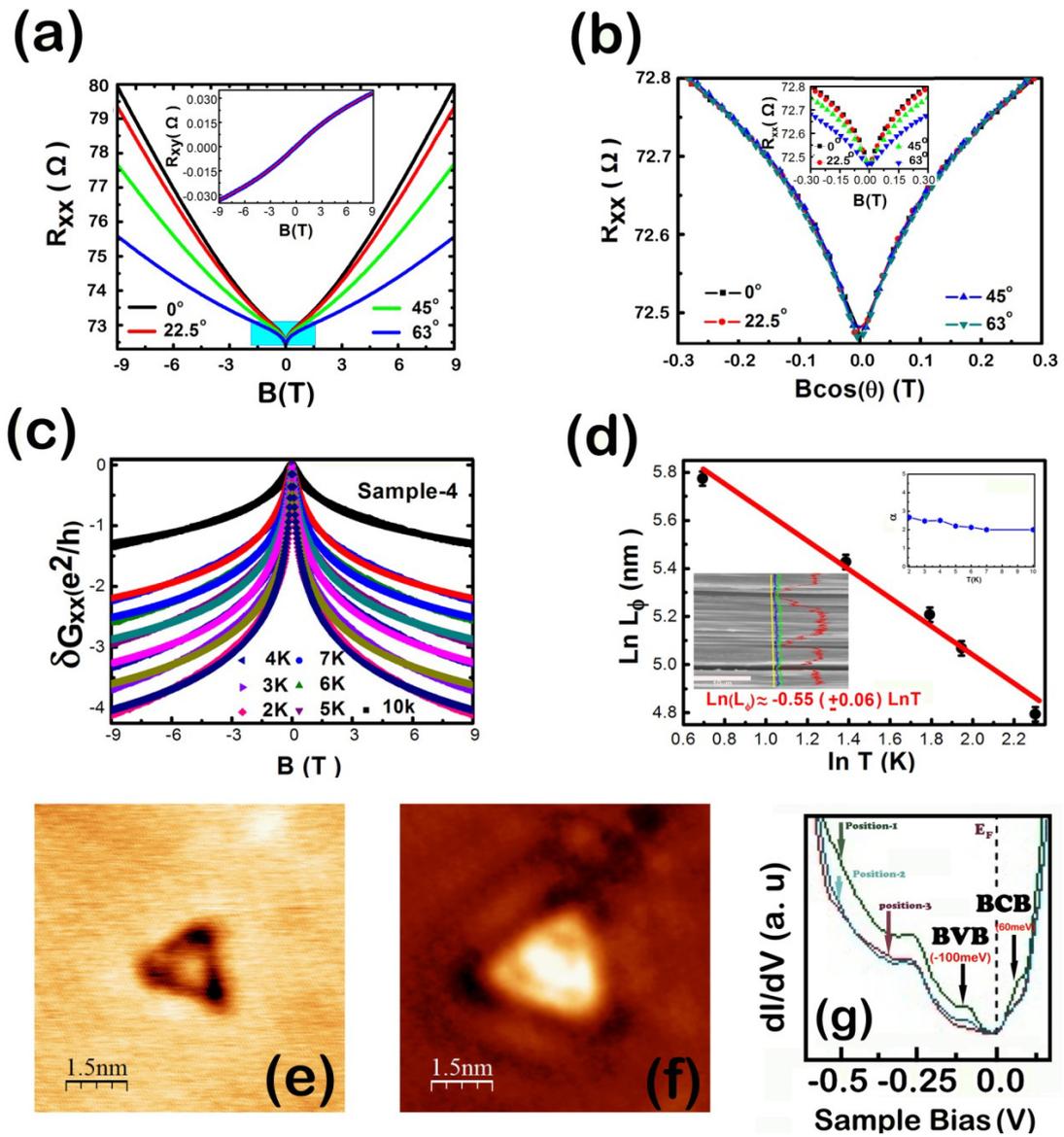

**Figure 2. Experimental evidence on the 2D Topological Surface Position In Sample-4.**

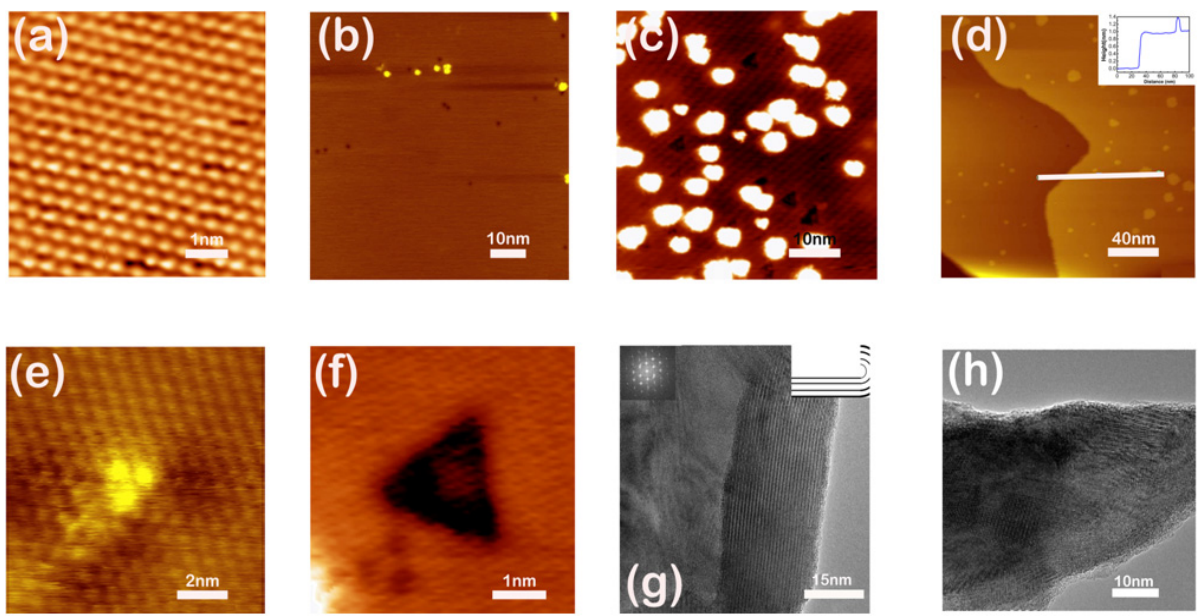

**Figure 3. Defects characterizations by STM and TEM.**

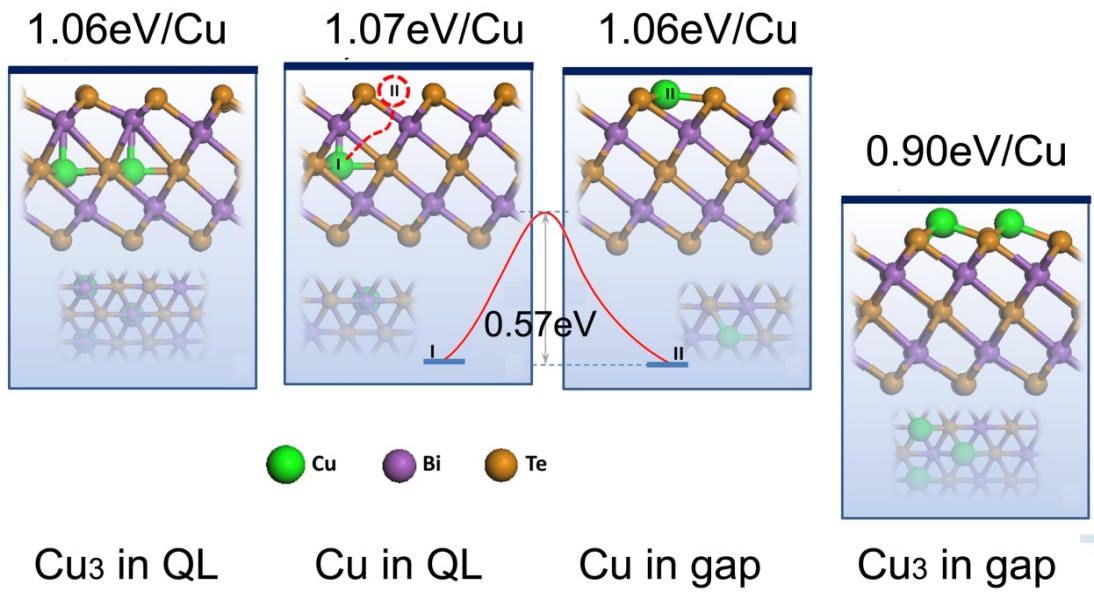

Figure 4. modeling the aging process.